\documentclass[%
 reprint,
superscriptaddress,
bibnotes,
 amsmath,amssymb,
 aps,
]{revtex4-1}

\usepackage{graphicx}
\usepackage{dcolumn}
\usepackage{bm}
\usepackage{xcolor}
\usepackage{commath}

\begin{document}

\title{Resistively-detected NMR lineshapes in a quasi-one dimensional electron system}

\author{M. H. Fauzi}
\affiliation{Department of Physics, Tohoku University, Sendai 980-8578, Japan}

\author{A. Singha}
\affiliation{Department of Electrical Engineering, Indian Institute of Technology Bombay, Mumbai 400076, India}

\author{M. F. Sahdan}
\affiliation{Department of Physics, Tohoku University, Sendai 980-8578, Japan}

\author{M. Takahashi}
\affiliation{Department of Physics, Tohoku University, Sendai 980-8578, Japan}

\author{K. Sato}
\affiliation{Department of Physics, Tohoku University, Sendai 980-8578, Japan}

\author{K. Nagase}
\affiliation{Department of Physics, Tohoku University, Sendai 980-8578, Japan}

\author{B. Muralidharan}
\affiliation{Department of Electrical Engineering, Indian Institute of Technology Bombay, Mumbai 400076, India}

\author{Y. Hirayama}
\affiliation{Department of Physics, Tohoku University, Sendai 980-8578, Japan}
\affiliation{Center for Spintronics Research Network, Tohoku University, Sendai 980-8577, Japan}

\date{\today}

\begin{abstract}

We observe variation in the resistively-detected nuclear magnetic resonance (RDNMR) lineshapes in quantum Hall breakdown. The breakdown is locally occurred in a gate-defined quantum point contact (QPC) region. Of particular interest is the observation of a dispersive lineshape occured when the bulk 2D electron gas (2DEG) is set to $\nu_{\rm{b}} = 2$ and the QPC filling factor to the vicinity of $\nu_{\rm{QPC}} = 1$, strikingly resemble the dispersive lineshape observed on a 2D quantum Hall state. This previously unobserved lineshape in a QPC points to simultaneous occurrence of two hyperfine-mediated spin flip-flop processes within the QPC. Those events give rise to two different sets of nuclei polarized in the opposite direction and positioned at a separate region with different degree of electronic spin polarization.

\end{abstract}

\pacs{73.43.Fj, 76.60.$-$k}

\maketitle

Recent advent in NMR technique through a resistive detection (RDNMR) has made it possible to study various spin physics in a 2D quantum Hall system\cite{Kumada, Zhang, MStern, Tiemann, Friess, Rhone, Piot}, and a quasi-1D channel\cite{Kou, Kawamura}. Despite the success achieved, a certain aspect related to the origin of the RDNMR lineshape variations noted experimentally in continuous wave (cw) mode is still poorly understood. One of them involved the puzzling observation of a dispersive lineshape in the quantum Hall state, a resistance dip followed by a resistance peak {resonance line} with increasing radio frequency\cite{Gervais_book}. It is first reported by Desrat et al\cite{Desrat2002} in the vicinity of $\nu_{\rm{b}} = 1$ and has been confirmed in a number of follow-up papers\cite{Piot, Tracy, Kodera, Dean, Bowers, Desrat2013, Desrat2015}. Similar dispersive like lineshape has been observed as well in the vicinity of $\nu_{\rm{b}} = 2/9$\cite{Gervais}, $\nu_{\rm{b}}=2/3$, $\nu_{\rm{b}}=1/3$\cite{Stern}, and at $\nu_{\rm{b}} = 2$ Landau level crossing\cite{Yang}. A number of appealing explanations has been put forward, but none of them provides a comprehensive explanation. Part of the reason why it still is difficult to unravel its physical origin is that we do not have a mature level of understanding about many-body 2D electronic states at the first Landau level yet, let alone their coupling to the nuclear spin. Thus, it would be highly desirable to study the lineshape variations in a platform where one can avoid such complexity.

\begin{figure}[t]
\begin{center}    
\centering
\includegraphics[width=\linewidth]{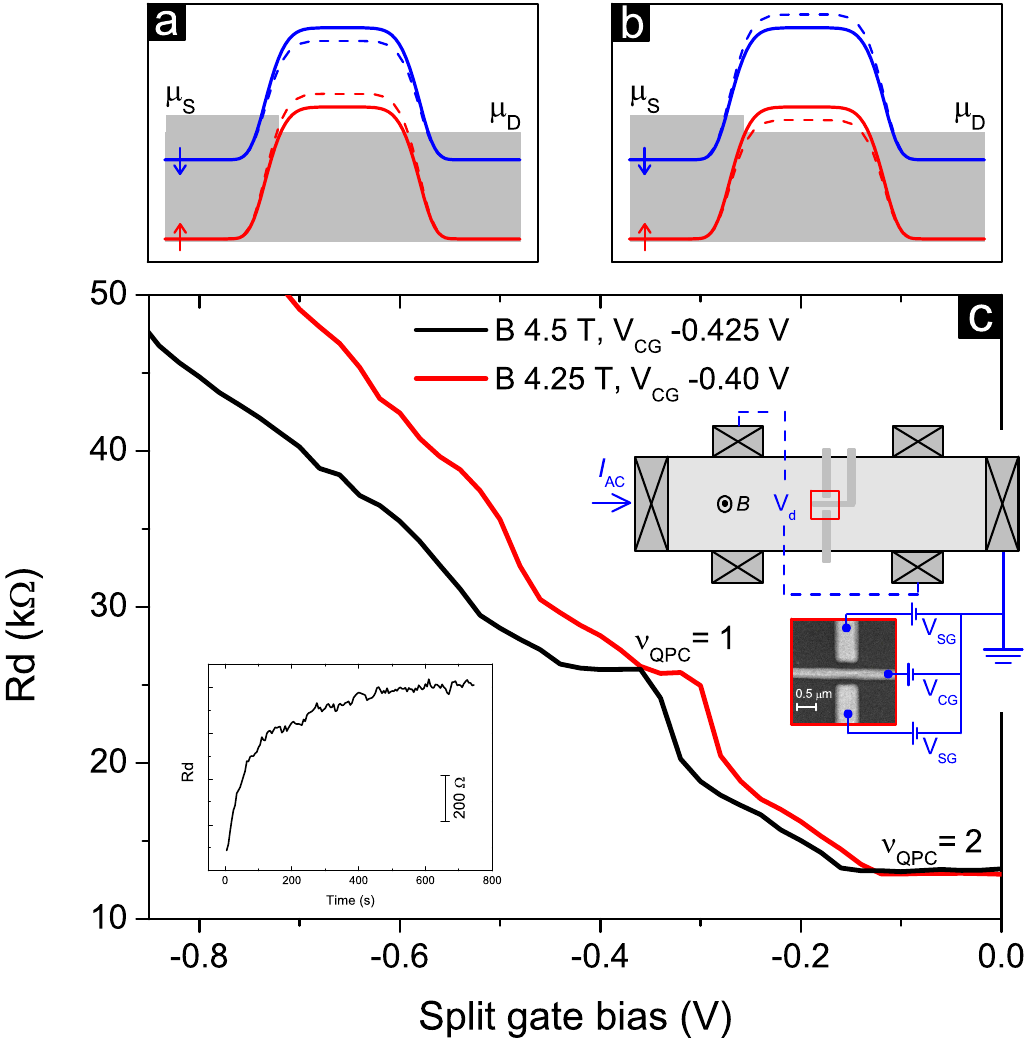}
\end{center}
\caption{{(a)-(b) Schematic of potential barrier seen by up-spin and down-spin electrons without (solid line) and with (dashed line) the presence of positive and negative nuclear polarization, respectively. The chemical potential window sits at $\nu_{\rm{QPC}} < 1$, so that only the up-spin channel affects the transport.} (c) Differential diagonal resistance $R_{\rm{d}} \equiv dV_{\rm{d}}/dI_{\rm{AC}}$ curve versus split gate bias voltage ($V_{\rm{SG}}$) at a field of $4.5$ T (black) and $4.25$ T (red). The left and right split gate are biased equally. {The center gate voltage $V_{\rm{CG}}$ is fixed to $-0.425$ V and $-0.4$ V, respectively}. Upper inset displays a schematic drawing of device. Cross marks represent Ohmic contact pads. Triple Schottky gates deposited on top of the Hall bar defined a quantum point contact (see SEM image). The lithographic gap(width) between(of) a pair of split gate is $600$($500$) nm. An extra gate (center gate) with lithographic width of $200$ nm is deposited in between the split gates. An excitation current $I_{\rm{AC}}=1$ nA with $f = 13.7$ Hz is applied to the device for transport measurement. Lower inset shows typical $R_{\rm{d}}$ time trace during current-induced dynamic nuclear polarization with $I_{\rm{AC}} = 10$ nA. }
\label{Fig01} 
\end{figure}

\begin{figure*}[t]
\begin{center}    
\centering
\includegraphics[width=\linewidth]{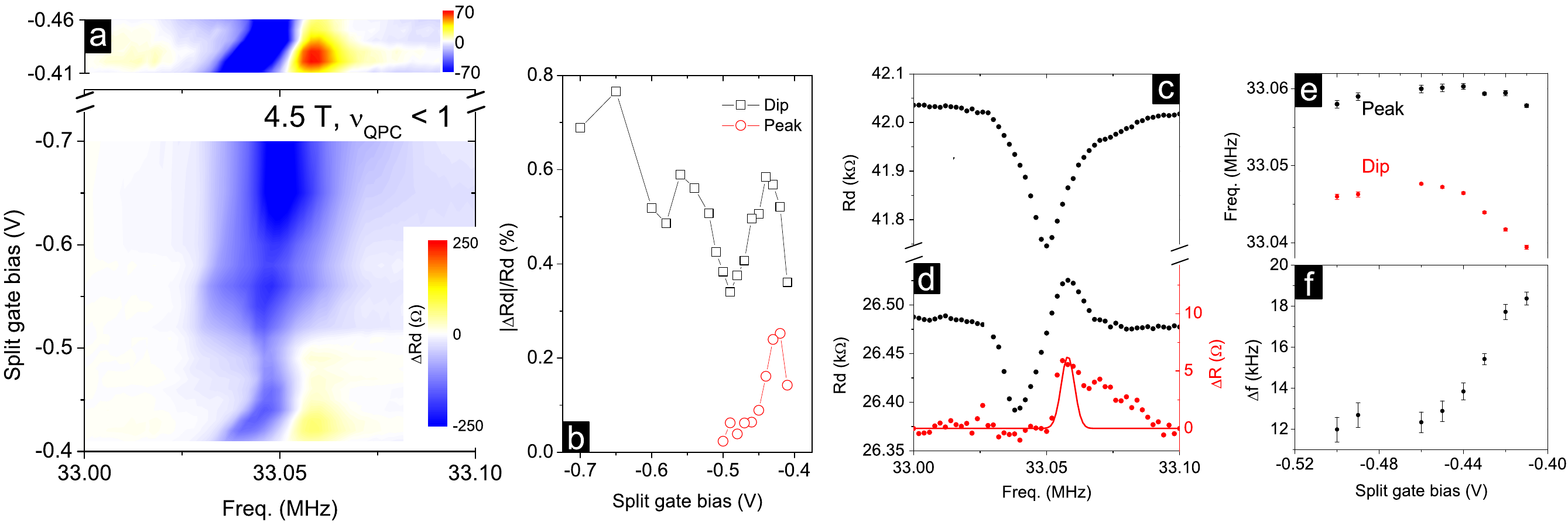}
\end{center} 
\caption{(a) Lower plot shows a 2D color map of $^{75}$As RDNMR traces at the upper flank of the $\nu_{\rm{QPC}} = 1$ plateau, $-0.70 \le V_{\rm{SG}} \le -0.41$ V, measured at $4.5$ T. The background resistance has been subtracted from the spectrum. Upper plot shows the blown-up spectra in between $-0.46 \le V_{\rm{SG}} \le -0.41$ V to accentuate the dispersive structure. (b) The RDNMR amplitude percentage vs split gate normalized to the off-resonance resistance, {$\abs{\Delta R{\rm{d}}}/R{\rm{d}}$}, extracted from panel (a). (c)-(d) Raw RDNMR data sliced at the $V_{\rm{SG}}=-0.7$ and $V_{\rm{SG}}=-0.41$ V, respectively. RDNMR in red dots superimposed in panel (d) measured very close to the bulk 2DEG $\nu = 2$ plateau, served as a reference signal with almost zero Knight shift. The signal is obtained by applying $I_{\rm{AC}} = 100$ nA. The red line is a Gaussian fit to the spectrum with the FWHM of 8.8 kHz. (e) The position of peak {resonance frequency} (black dots) and dip {resonance frequency} (red dots) extracted from panel (a) for $-0.50 \le V_{\rm{SG}} \le -0.41$ V. (f) The peak-to-dip {resonance frequency separation} ${\Delta}f$ extracted from panel (e). All the spectra measured with $I_{\rm{AC}} = 10$ nA (except the ref. signal) and RF power is $-30$ dBm.}
\label{Fig02} 
\end{figure*}

In this {Rapid Communication, we resort to a quasi-one dimensional system in a gate-defined quantum point contact (QPC) to study} various possible lineshapes including the dispersive lineshape noted experimentally in cw mode. Unlike on the 2D system, the mechanism for generation and resistive detection of nuclear spin polarization is tractable, allowing conveniently a direct interpretation of the observed lineshapes. 

{Generation and detection of nuclear spin polarization are achieved by setting the filling factor in the bulk 2DEG to $\nu_{\rm{b}} = 2$ and $\nu_{\rm{QPC}} = 1$ in the QPC\cite{Wald, Dixon, Machida, edge1, edge2, edge3, Masubuchi, Corcoles, Keane, Chida}. Fig. \ref{Fig01}(a)-(b) schematically displays how the nuclear polarization affects the transmission probability through the potential barrier of the QPC. For $\nu_{\rm{QPC}} < 1$ (the down-spin channel $T_{\downarrow}$ does not affect the transport), the up-spin channel $T_{\uparrow}$ sees an increase(decrease) in the barrier potential in the presence of positive(negative) nuclear polarization, where positive (negative) means nuclear polarization is parallel (opposite) to the external magnetic field. Consequently, the transmission probability of the up-spin channel is reduced(enhanced). Therefore, the transmission is modified by a dynamic nuclear polarization (DNP) under a steady state where nuclear spins diffuse from the polarized regions to the center of the QPC. At sufficiently high current densities, there are two possible tractable DNPs by hyperfine-mediated inter edge spin-flip scattering within the lowest Landau level, namely forward and backward spin-flip scatterings\cite{Wald, Dixon, Aniket}. The first (second) one involves a spin-flip scattering from the forward propagating up-spin (down-spin) channel to the forward (backward) propagating down-spin (up-spin) channel, which in turn produces the positive (negative) nuclear polarization through the spin flip-flop process. On sweeping the rf field after the polarization reaches a steady state, those two different sets of nuclear polarization would leave a different trace in the RDNMR signal; with the positive (negative) one resulting in a resistance dip (peak). Here we demonstrate that under certain electronic state in the QPC, those two sets of nuclei can be generated simultaneously in a separate region within the QPC. Since they experience different degree of electron spin polarization, one can observe a combination of a resistance dip and peak resonance line in the RDNMR spectrum, namely dispersive lineshape.}

Our studies are carried out on a 20-nm-wide doped GaAs quantum well with the 2DEG located $165$ nm beneath the surface. The wafer is photo-lithographically carved into a $30$-$\mu$m-wide and $100$-$\mu$m-long Hall bar geometry. The low temperature electron mobility is $84.5$ m$^2$/Vs at an electron density of $1.0 \times 10^{15}$ m$^{-2}$. A single QPC defined by triple Schottky gates is patterned on top of the Hall bar by Ti/Au evaporation.  The bulk 2DEG density $n$ can be tuned by applying back gate voltage ($V_{\rm{BG}}$) to Si-doped GaAs substrate. It enables us to control the filling factor of interest in the bulk 2DEG $\nu = \frac{h}{eB}n$ with back gate $V_{\rm{BG}}$ and magnetic field $B$. The samples are mounted inside a single-shot cryogenic-free $^3$He refrigerator with a temperature of 300 mK. A six-turn rf coil wrapped the sample to be able to apply an oscillating magnetic field in the plane of the 2DEG. Throughout this study, the amplitude of rf power delivered to the top of the cryostat is fixed to $-30$ dBm (unless specified otherwise).

{Fig. \ref{Fig01}(c)} displays two sets of diagonal resistance traces as a function of split gate bias voltage across the QPC measured at a field of $4.5$ (black line) and $4.25$ (red line) T. {The center gate is fixed to $V_{\rm{CG}} = -0.425$ V and $V_{\rm{CG}} = -0.4$ V, respectively\cite{Misc}}. We start with fully filled first Landau level in the bulk 2DEG ($\nu_{\rm{b}}=2$), where both the up-spin and down-spin electrons are available for transmission. Applying negative voltage on the split gates allows us to selectively transmit the up-spin channel through the constriction and reflect the down-spin channel. The nuclear spins is dynamically polarized by applying $I_{\rm{AC}} = 10$ nA at a selected operating point along the diagonal resistance trace on both sides of the $\nu_{\rm{QPC}} = 1$ plateau. Typically, the resistance increases exponentially and reaches a point of saturation on the time scale of a few hundred seconds with the characteristic exponential rise time of about $150$ seconds (see the lower inset of Fig. \ref{Fig01}), similar time scale characteristic is reported previously on other QPC structures\cite{Corcoles}. Once the resistance saturated, the rf is swept across the Larmor frequency of $^{75}$As nuclei while measuring its resistance. The rf sweep rate is set to $100$ Hz/s\cite{Supp3}.

\begin{figure}[t]
\begin{center}    
\centering
\includegraphics[width=\linewidth]{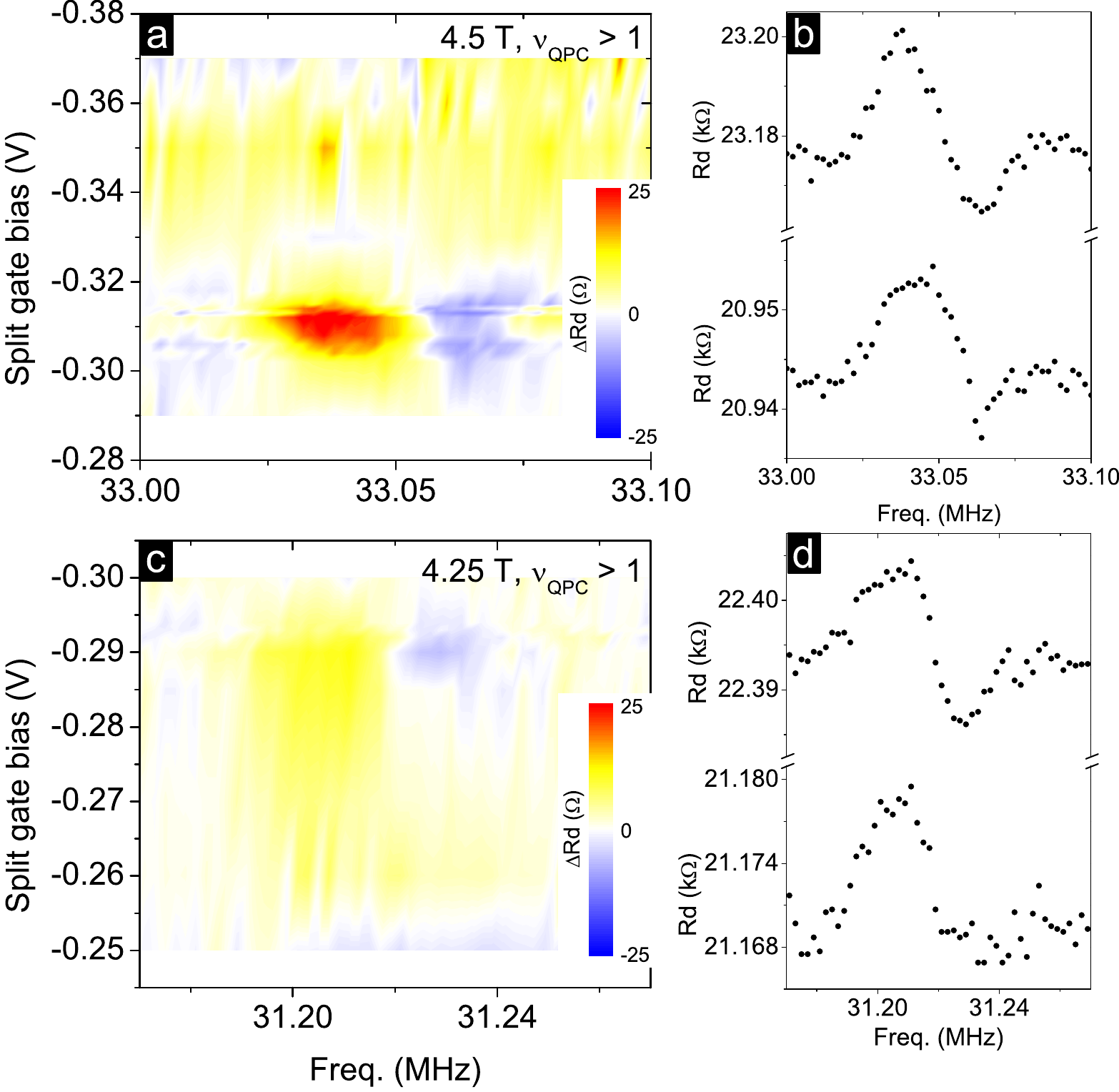}
\end{center} 
\caption{ (a) 2D color map of $^{75}$As RDNMR traces at the lower flank of the $\nu_{\rm{QPC}} = 1$ plateau ($\nu_{\rm{QPC}} > 1$) measured at a field of $4.5$ T. (b) Raw RDNMR traces sliced at $V_{\rm{SG}} = -0.313$ (upper) and $V_{\rm{SG}} = -0.302$ (lower) V, respectively. (c) 2D color map of $^{75}$As RDNMR traces at the lower flank of the $\nu_{\rm{QPC}} = 1$ plateau ($\nu_{\rm{QPC}} > 1$) measured at a field of $4.25$ T. (d) Raw RDNMR traces sliced at $V_{\rm{SG}} = -0.29$ (upper) and $V_{\rm{SG}} = -0.285$ (lower) V, respectively.}
\label{Fig03} 
\end{figure}

We observe variation in the RDNMR lineshape spectra on both flank of the $\nu_{\rm{QPC}} = 1$ plateau as displayed in Fig. \ref{Fig02} and \ref{Fig03}. Let us start with the RDNMR spectra for $\nu_{\rm{QPC}} < 1$ case observed at a field of $4.5$ T shown in Fig. \ref{Fig02}(a), measured from $V_{\rm{SG}} = -0.41$ up to $V_{\rm{SG}} = -0.7$ V. For ease of comparison, we plot the resistance variation $\Delta R_{\rm{d}}$ with respect to the off-resonance resistance at $f = 33$ MHz. The salient feature appears in a narrow portion of the split gate bias voltage region, $-0.50 \le V_{\rm{SG}} \le -0.41$ V, very close to the $\nu_{\rm{QPC}} = 1$ plateau. The spectra have a curious dispersive lineshape, strikingly resemble the dispersive lineshape previously observed in a number of reports on a 2D quantum Hall system in the vicinity of $\nu_{\rm{b}} = 1$\cite{Desrat2002, Piot, Tracy, Kodera, Dean, Bowers, Desrat2013, Desrat2015}. The lineshape we observe in our system is found to be highly sensitive to the rf power such that the resistance peak {resonance line} vanishes at a relatively high rf power of -15 dBm\cite{Supp4}.

The corresponding signal amplitude normalized to the off resonance resistance {$\abs{\Delta R_{\rm{d}}}/R_{\rm{d}}$} is displayed in Fig. \ref{Fig02}(b). All the signal amplitude observed here falls below $1\%$, similar to the previous reports in Ref. \cite{Corcoles, Keane}. Starting from the observable signal closest to the plateau $V_{\rm{SG}} = -0.41$ V, the dip amplitude shows a sharp upturn and reaches a maximum value at $V_{\rm{SG}} = -0.44$ V. It is then followed by a downturn and takes on a minimum value at $V_{\rm{SG}} = -0.50$ V, precisely at the transition between dispersive-to-single lineshape. The peak amplitude has a smaller amplitude than the dip amplitude and shows a monotonically decrease from $V_{\rm{SG}} = -0.42$ V and eventually vanishes at $V_{\rm{SG}} = -0.51$ V. The spectrum evolves into an expected single dip lineshape for $V_{\rm{SG}} \le -0.51$ V with the signal amplitude gradually increases. It can be partially explained by an increase in the current density locally in the constriction. Altogether, the facts that the lineshapes, signal amplitudes, as well as resonance point variations with the split gate bias voltage constitute firm evidence that the nuclei is polarized locally in the QPC.

We plot in Fig. \ref{Fig02}(c)-(d) the raw RDNMR spectra at the two most extreme cases $V_{\rm{SG}} = -0.70$ and $V_{\rm{SG}} = -0.41$, respectively. In order to extract the Knight shift for each spectrum, here {we plot in} Fig. 2(d) (red dots) the reference signal taken close to $\nu_{\rm{b}} = 2$ with nearly zero Knight shift. The spectrum is fitted with a Gaussian function\cite{Masubuchi}, centered at $33.057$ MHz and FWHM of $8.8$ kHz (red line). Note that the long tail in the higher radio frequency side in the reference spectrum is nothing but reflects a long T1 time\cite{Hashimoto}. Comparing with the reference signal, the observed spectrum at $V_{\rm{SG}} = -0.70$ V is only Knight shifted by about $8$ kHz, reasonable value for the spectrum very far from the plateau at a field of $4.5$ T. The dip frequency in the dispersive lineshape at $V_{\rm{SG}} = -0.41$ V gives the largest observable shift by about $18$ kHz. Interestingly, its peak frequency appears to be substantially unshifted as it is aligned reasonably well with the reference resonance point. RDNMR measurement performed at a smaller field of $4.25$ T reveals similar lineshape patterns\cite{Supp6}. 

Fig. \ref{Fig02}(e) displays the dip and peak resonance line points extracted from the split gate bias voltage segment between $-0.41$ to $-0.50$ V, where the dispersive lineshape is observed. The peak resonance line lies at the resonance reference point with very small variation throughout the range, substantially not Knight shifted. On the other hand, the dip resonance line is upshifted in a linear fashion up to $V_{\rm{SG}} = -0.46$ V and then followed by a slight downshift. The resulting $\Delta f$ values extracted from panel (e) is plotted in Fig. 2(f). The $\Delta f$ value continuously drops down to $12$ kHz in an obviously linear fashion up until $V_{\rm{SG}} = -0.46$ V from its initial value of $18.3$ kHz at $V_{\rm{SG}} = -0.41$ V, bearing a similarity to $\Delta f - B$ plot around $\nu_{\rm{b}}=1$ observed on the 2D system\cite{Desrat2013}. The value remains constant at about $12$ kHz throughout the remaining split gate values, an indication that the electronic state in the QPC does not change significantly. Similar trend is observed as well for a field of $4.25$ T\cite{Supp7}.

We now move on to discuss the RDNMR taken at the opposite side of the plateau ($\nu_{\rm{QPC}} > 1$) as shown in Fig. \ref{Fig03}. The data show similar lineshape trend, but with inverted signal and much smaller amplitude than its counterpart. At a field of $4.5$ T displayed in Fig. \ref{Fig03}(a), the RDNMR signal is visible only in a confined split gate bias range, $ -0.32 \le V_{\rm{SG}} \le -0.30 $ V. The spectra measured very close to the plateau are hindered by a large resistance fluctuation in particular at the point where the diagonal resistance abruptly changes. Nevertheless, one can verify the existence of the inverted dispersive lineshape for $\nu_{\rm{QPC}} > 1$ (see the line-cuts at $V_{\rm{SG}}=-0.313$ and $V_{\rm{SG}}=-0.302$ V in Fig. \ref{Fig03}(b) for better visual). The RDNMR signal measured at a field of $4.25$ T displayed in Fig. \ref{Fig03}(c) has less resistance fluctuation and hence offers better signal to noise ratio. The inverted dispersive lineshape appears at $V_{\rm{SG}}=-0.29$ V (upper Fig. \ref{Fig03}d) and turns into a resistance peak lineshape at $V_{\rm{SG}}=-0.285$ V (lower Fig. \ref{Fig03}d). In contrast to the case for  $\nu_{\rm{QPC}} < 1$ where the RDNMR signal is observed in a wide range of split gate bias voltages, the signal observed here vanishes very quickly far from the $\nu_{\rm{QPC}} = 1$ plateau region. Recall that the hyperfine-mediated spin flip-flop process relies on the spatial overlap between the up-spin and down-spin channels\cite{Wald}. Thus, the absence of RDNMR signal indicates the critical current for breakdown is higher than for $\nu_{\rm{QPC}} < 1$ since the channel is opened wider\cite{Hwang}.

\begin{figure}[t]
\begin{center}    
\centering
\includegraphics[width=\linewidth]{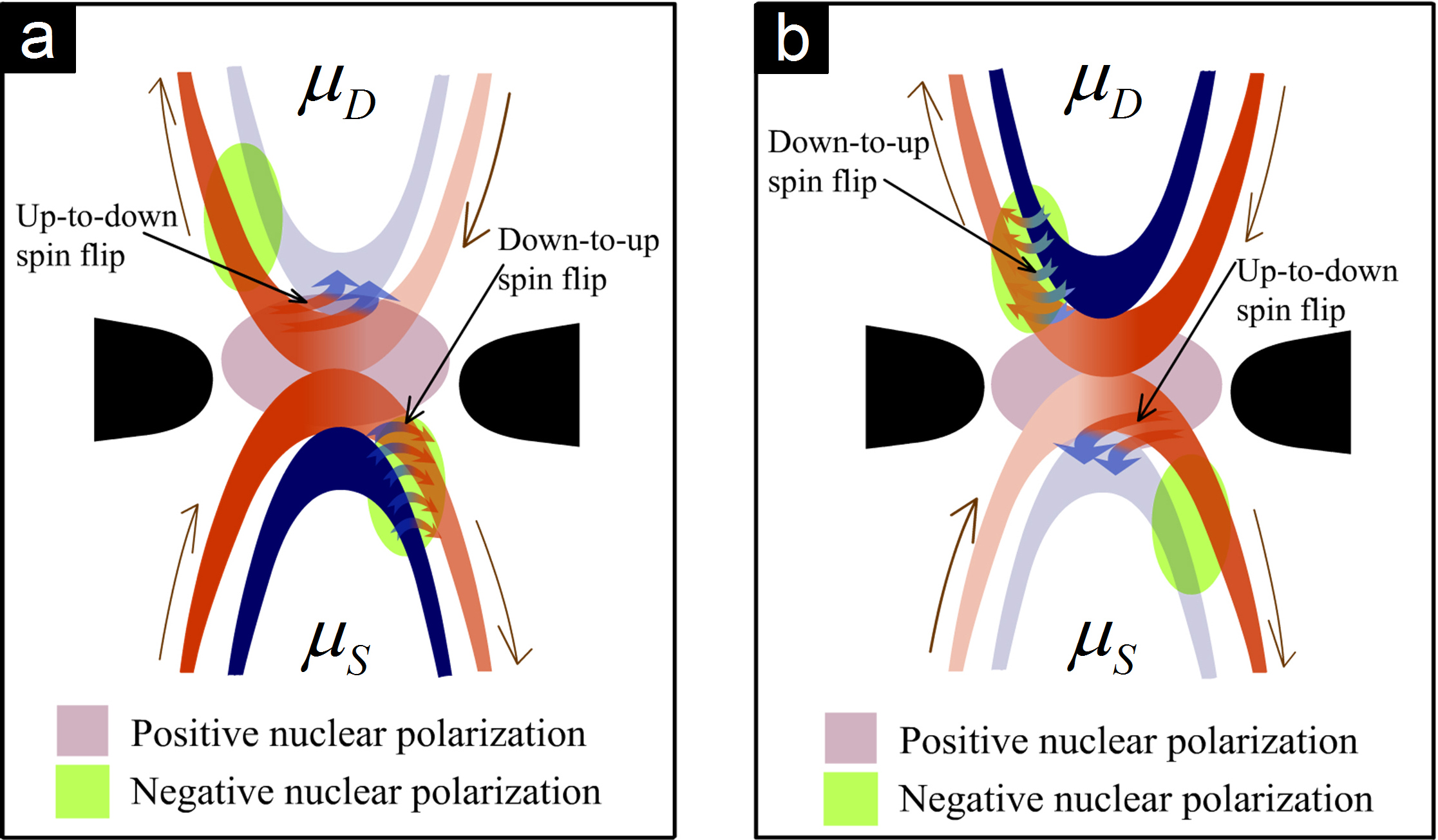}
\end{center} 
\caption{(a) Schematics of Landauer-B\"uttiker edge channel with forward (up-to-down spin flip) and backward (down-to-up spin flip) hyperfine-mediated spin-flip scatterings occurred at filling factor slightly smaller than $\nu_{\rm{QPC}} \approx 1$ during the first half-clock alternating current cycle ($\mu_S > \mu_D$) and (b) during the second half-clock alternating current cycle ($\mu_S < \mu_D$). Lighter edges indicate an empty channel while darker edges indicate a filled channel. The drain is held at ground ($\mu_{D}=0$) while the source chemical potential $\mu_{S}=0$ oscillates at a frequency of $13.7$ Hz.}
\label{Fig04} 
\end{figure}

The results presented in Fig. \ref{Fig02}$-$\ref{Fig03} provide important insights onto mechanisms leading to the dispersive lineshape observed in the vicinity of $\nu_{\rm{QPC}} = 1$ plateau. {Fig. \ref{Fig04} displays all possible hyperfine-mediated spin-flip scattering events where the QPC filling factor is tuned slightly less than 1 for two different alternating current cycles.} {The forward and backward spin-flip scattering could occur simultaneously within the QPC. The forward scattering occurs at the central region of the QPC where the degree of electron spin polarization is finite, not zero. On the other hand, the backward spin-flip scattering occurs slightly outside the central region where the electron spin polarization is zero. Those scattering events polarize the nuclei in opposite direction and spatially separated. On sweeping of rf with increasing frequency, the positive nuclear polarization is destroyed first due to Knight shift. It results in an increase in the transmissivity of the up-spin channel. On further sweeping the rf, the positive nuclear polarization starts to build up and negative nuclear polarization is destroyed. This results in a decrease in the transmissivity of the up-spin channel. The backward spin-flip scattering is highly suppressed when the QPC filling factor is further tuned to $\nu_{\rm{QPC}} < 1$, leaving only positive nuclear polarization build-up at the central region of the QPC. The RDNMR spectrum switches from dispersive-like to dip resonance lineshape.} {In this scenario, the Knight shift 
at the central region is determined by $K_S \propto \left(n_{\uparrow} - n_{\downarrow}\right) \propto \left(T_{\uparrow} - T_{\downarrow}\right)$, where $n_{\uparrow}(n_{\downarrow})$ and $T_{\uparrow}(T_{\downarrow})$ are up(down)-spin 
electron density and up(down)-spin transmission probability, respectively. The Knight shift reaches a maximum value when the up-spin channel is completely transmitted ($T_{\uparrow}=1$) while the down spin channel is completely reflected ($T_{\downarrow}=0$). It decreases with reduction of $T_{\uparrow}$, agreeing well with the experimental data shown in Fig. \ref{Fig02}(e).}

{For $\nu_{\rm{QPC}} > 1$ case, similar scenario happens. However, the Overhauser field from the polarized nuclei now affects the transmission of the down-spin channel while the fully transmitted up-spin channel is left unaffected. The nuclear polarization influences the transmissivity of the down-spin channel in an opposite way than that of the up-spin channel. This is the reason why the RDNMR spectrum gets inverted as experimentally confirmed in Fig. \ref{Fig03} and noted in Ref. \cite{Keane}.}

To summarize, here we observe four variation of the RDNMR lineshapes in a gate-defined QPC. Of particular interest is the emergence of the dispersive lineshape in the RDNMR signal when the bulk filling factor is set to $\nu_{\rm{b}} = 2$ and the QPC filling factor to the vicinity of the $\nu_{\rm{QPC}} = 1$ plateau. It can be accounted by considering simultaneous occurrence of two hyperfine-mediated spin-flip scattering events due to current-induced dynamic nuclear polarization. These phenomena give rise to localized regions with opposite nuclear polarization in the QPC. Although both of them are in contact with electrons in the QPC, they polarize in a region with different degree of electron spin polarization. Our experimental results further cemented the idea that the observation of the dispersive lineshapes on the 2D system, in particular around $\nu_{\rm{b}}=1$, should reflect the nuclear spin interaction with two electronic sub-systems as suggested by the authors in Ref.\cite{Piot, Desrat2013}. 

We would like to thank K. Muraki of NTT Basic Research
Laboratories for supplying high quality wafers for this study. We thank K. Hashimoto, K. Akiba, T. Aono, T. Tomimatsu, B. Friess, A. Micholic, and D. G. Austing for helpful discussions and/or technical assistance. M.H.F. and Y.H. acknowledge support from Multi-Dimensional program, Tohoku University. A. S., M.T., K.N., and Y.H. acknowledge support from Graduate Program in Spintronics, Tohoku University. B. M. and Y. H. acknowledge support from WPI-AIMR, Tohoku University. Y.H. acknowledges financial support from KAKENHI Grants Nos. $26287059$ and $15\rm{H}05867$. B. M. and A. S. acknowledge funding from the Department of Science and Technology, India under the Science and Engineering Board (SERB) grant no. SERB/F/3370/2013-2014.


\begin{thebibliography}{35}
\bibitem [{Mis()}]{Misc}%
  \BibitemOpen
  \href@noop {} {}\bibinfo {note} {Tuning the center gate bias to $-0.425$($-0.4$) V for a field of $4.5$($4.25$) T, we effectively make the channel wide so that the resistance increases gradually for $\nu_{\rm{QPC}} < 1$. This gradual changes in the diagonal resistance make systematic RDNMR measurement possible for $\nu_{\rm{QPC}} < 1$.}\BibitemShut {Stop}%
\end{thebibliography}
%

\end{document}